\documentclass[aps,prb, twocolumn,showpacs,preprintnumbers,superscriptaddress]{revtex4}

\usepackage[dvipdfm]{graphicx}% Include figure files
\usepackage{dcolumn}% Align table columns on decimal point
\usepackage{bm}% bold math

\begin{document}


\title{Effect of pressure on magneto-transport properties in the superconducting and normal phases of the  metallic double chain compound Pr$_{2}$Ba$_{4}$Cu$_{7}$O$_{15-\delta }$}

\author{Masayoshi Kuwabara}
\author{Michiaki Matsukawa} 

\email{matsukawa@iwate-u.ac.jp }
%\thanks{}

\author{Keisuke Sugawara}
\author{Haruka Taniguchi} 
%\author{Junichi Echigoya} 
\affiliation{Department of Materials Science and Engineering, Iwate University, Morioka 020-8551, Japan}
\author{Akiyuki  Matsushita}
\affiliation{National Institute for Materials Science, Ibaraki 305-0047}
\author{Makoto Hagiwara}
%\author{Tsuyoshi Miyazaki}
\affiliation{Kyoto Institute of Technology, Kyoto 606-8585,Japan}
\author{Kazuhiro Sano}
\affiliation{Department of Physics Engineering, Mie University, Tsu 514-8507,Japan}
\author{Yoshiaki Ono}
%$\author{Yuh Yamada}
\affiliation{Department of Physics, Niigata University, Niigata 950-2181,Japan}
\author{Takahiko Sasaki}
%\author{Yuichiro Hayasaka}
\affiliation{Institute for Materials Research, Tohoku University, Sendai 980-8577,Japan}

%\author{Junichi Echigoya$^{1}$}

\date{\today}

\begin{abstract}
To examine the electronic phase diagram  of superconducting CuO double chains, we report the effect of external pressure on the magneto-transport properties   in  superconducting and non-superconducting  polycrystalline samples of Pr$_{2}$Ba$_{4}$Cu$_{7}$O$_{15-\delta }$ at low temperatures (1.8-40 K) under various magnetic fields (up to 14 T).  
In the as-sintered non-superconducting sample,  the magneto-resistance (MR)  follows a power law of  $H^{3/2}$ at low temperatures, which is in no agreement with the  $H^{2}$ dependence of MR in the PrBa$_{2}$Cu$_{4}$O$_{8}$ system.
The negative pressure dependence of the superconducting phase is qualitatively consistent with a theoretical prediction on the basis of the Tomonaga-Luttinger Liquid theory. 
The 48-h-reduced superconducting sample at ambient pressure exhibits  no clear increase in MR for $T > T_{c,on}=26.5$ K. 
In contrast, with the application of pressure to the superconducting  sample, the MR  effects reappear and are also well fitted by  $H^{3/2}$.
The model of slightly warped Fermi surfaces explains  not only the MR effect of the non-superconducting sample, but is also related to the reasons for the pressure-induced MR phenomena of the superconducting sample. 

%These anomalous behaviors in  the normal state  were discussed in the view point of topologies of Fermi surfaces as theoretically predicted, which are varied from warped  to  straight form  upon increasing the carrier content.  
%Furthermore, the suppression of the superconducting phase due to the application of pressure was explained by the enhanced one-dimensionality of CuO double chains. 

\end{abstract}

%\keywords{}
\renewcommand{\figurename}{Fig.}
\maketitle

\section{INTRODUCTION}

Since the discovery of high-$T_\mathrm{c}$ copper-oxide superconductors,  researches have focused on the unconventional superconductivity 
on the two-dimensional CuO$_{2}$ planes in some cuprates. 
In quasi one-dimensional (1D) ladder system without CuO$_{2}$ planes, it is well known that the superconductivity at  $T_\mathrm{c}$= 12 K  appears only under the application of high pressure.\cite{UE96}
In our previous study of Pr-based cuprates with metallic CuO double chains and insulating CuO$_{2}$ planes, Pr$_{2}$Ba$_{4}$Cu$_{7}$O$_{15-\delta}$  is found to be a new superconductor  with a higher $T_\mathrm{c}$ (15 K) after a reduction treatment.\cite{MA04}
A nuclear quadrupole resonance (NQR) study has revealed that the newly discovered superconductivity is realized at the CuO double-chain block. \cite{WA05}

%strongly correlated electron systems  have been extensively investigated. Besides the physical properties 
%of two-dimensional CuO$_{2}$ planes, researches have focused on the physical role of one-dimensional (1D) CuO chains
%in some families of high-$T_\mathrm{c}$ copper oxides.

Structurally, the Pr-based cuprates, PrBa$_{2}$Cu$_{3}$O$_{7-\delta}$ (Pr123) and PrBa$_{2}$Cu$_{4}$O$_{8}$ (Pr124), are identical to their corresponding
Y-based high-$T_\mathrm{c}$  superconductors, YBa$_{2}$Cu$_{3}$O$_{7-\delta}$ (Y123) and  YBa$_{2}$Cu$_{4}$O$_{8}$ (Y124).
Pr123 and Pr124 compounds have insulating CuO$_{2}$ planes and are non-superconductive. \cite{SO87,HO98}  
The suppression of superconductivity in the Pr substitutes has been explained by the hybridization of 
Pr-4$f$ and O-2$p$ orbitals.\cite{FE93}  
The crystal structure of Pr124 with CuO double chains differs from that of Pr123 with CuO single chains. 
It is well known that CuO single chains in Pr123 and CuO double chains in Pr124 show semiconducting and metallic behaviors, respectively.\cite{MI00}
The carrier concentration of doped double chains of Pr124 is difficult to vary,  because it is thermally stable up to high temperatures.
%In addition, Pr ions in both Pr123 and Pr124 become antiferromagnetic ordered at the antiferromagnetic transition temperature $T_\mathrm{N}=17$ K.\cite{Li89,YA97} 

The compound Pr$_{2}$Ba$_{4}$Cu$_{7}$O$_{15-\delta}$ (Pr247) is an  intermediate between  Pr123 and Pr124.   
In this compound,  CuO single-chain and double-chain blocks are alternately stacked along the $c$-axis such as \{-D-S-D-S-\} sequence \cite{BO88,YA94} (see Fig.\ref{Xray}). 
Here, S and D denote CuO single-chain and double-chain blocks along the $b$-axis, respectively.
The physical properties of  the metallic CuO double chains can be examined 
by controlling the oxygen content along the semiconducting CuO single chains.
Anisotropic resistivity measurements of single-crystal Pr124 have revealed that metallic transport arises by the conduction along the CuO double chains.\cite{HO00}
In oxygen removed polycrystalline Pr$_{2}$Ba$_{4}$Cu$_{7}$O$_{15-\delta }$, the superconductivity appears at an onset temperature $T_\mathrm{c,on}$ of $ \sim $15 K. \cite{MA04} 
Hall coefficient measurements of  superconducting Pr247 with $T_\mathrm{c,on}=15$ K have revealed that at intermediate temperatures below 120 K, the main carriers change from holes to electrons, as the temperature decreases. Accordingly, this compound is an electron-doped superconductor.\cite{MA07} 
In our previous study, we examined the effect of magnetic fields on the superconducting phase of Pr247.\cite{CH13}  Despite of the resistive drop associated with the superconducting transition, we found that the diamagnetic signal was strongly suppressed as expected in  the 1D superconductivity of  CuO double chains.  We also reported the temperature dependence of the Hall coefficient in superconducting Pr247 with a higher $T_\mathrm{c,on}\sim 27$ K.\cite{TA13}
Our findings indicated that the superconducting transition temperature increased because  the density of doped electron carriers became denser under the reduction treatment, consistent with a theoretical prediction.\cite{SA05}

In this paper, we demonstrate the magneto-transport properties of the electron-doped metallic double-chain compound Pr$_{2}$Ba$_{4}$Cu$_{7}$O$_{15-\delta }$, which has a  higher $T_\mathrm{c,on}$ (26.5 K), under the application of different hydrostatic pressures (up to 1.6 GPa). 
%Figure \ref{TEM} displays the typical crystal structure of Pr$_{2}$Ba$_{4}$Cu$_{7}$O$_{15-\delta }$, in which the CuO metallic double chains and semiconducting single chains are  alternately stacked  along the $c$-axis. 
Section II outlines the experimental methods, and Sec. III presents the magneto-transport  properties  of superconducting and non-superconducting polycrystalline Pr247 samples under external pressures. 
%Along with the magneto-transport data, we report the effect on pressure on the magneto-resistance in the as-sintered non-superconducting and reduced superconducting samples of Pr247. 
These results  are discussed  in the view point of topologies of Fermi surfaces, which are varied from warped  to  straight form  upon increasing the carrier content.  
%These results  are discussed in terms of a three-level system (quasi triplet), in which the ground-state of  the Pr$^{3+}$ ions is split by the crystal field effect.  
The final section is devoted to a summary.

\section{EXPERIMENT}
\begin{figure}[ht]
\includegraphics[width=10cm]{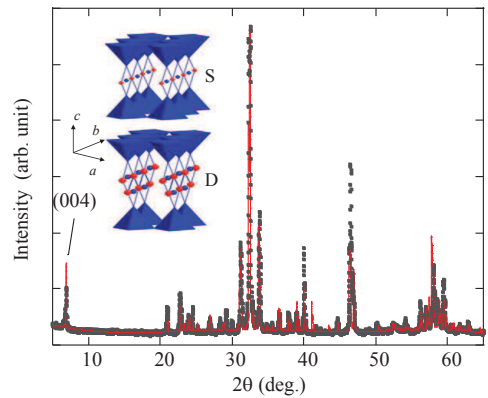}
\caption{(color online) X-ray diffraction patterns of as-sintered polycrystalline  Pr$_{2}$Ba$_{4}$Cu$_{7}$O$_{15-\delta }$. The (004) peak corresponds to one of typical Miller indexes of Pr247.  The calculated curve is obtained using the lattice parameters in the text. 
Inset shows  the crystal structure of Pr247. S and D denote CuO single-chain and double-chain blocks along the $b$-axis, respectively.}
\label{Xray}
\end{figure}

Polycrystalline samples of Pr$_{2}$Ba$_{4}$Cu$_{7}$O$_{15-\delta }$(Pr247) were synthesized by the citrate pyrolysis method.\cite{HA06} After several annealing processes,  the resulting precursors were pressed into  a pellet and  calcined at 875-887 $ ^{ \circ }$C for an extended period over 120-180 h under ambient oxygen pressure.
The oxygen in the as-sintered sample was removed by reduction treatment   in a vacuum  at 500 $ ^{ \circ }$C for 48 h,  yielding  a superconducting material. 
Typical dimensions of the pelletized rectangular sample were $9.9\times 2.4\times 1.7$ mm$^{3}$.
X-ray diffraction data  revealed that the as-sintered polycrystalline samples are an almost single phase with an orthorhombic structure ($Ammm$), as shown in Fig.\ref{Xray}.  
The lattice parameters of the as-sintered sample  are $ a=3.8753$ \AA, $ b=3.9107$ \AA, and $ c=50.713$ \AA,  which are in fair agreement with those obtained by a previous study\cite{YA05}. 
In a previous  TEM image of superconducting Pr247, we observed a regular stacking structure of \{-D-S-D-S-\} along the $c$-axis direction\cite{KO16}. 
%The Pr247 sample with $T_\mathrm{c}^\mathrm{on}=26.5$ K was observed by high-resolution transmission electron microscopy  using a JEOL3010 microscope operated at 300 kV at Tohoku University.
%The local crystal structure was analyzed from images  obtained by the high-angle annular dark-field scanning transmission electron microscope  method.\cite{PE90} 
The oxygen deficiency in the sample with $T_\mathrm{c,on}=26.5$ K  prepared by the citrate method was estimated to be  $\delta = 0.56$  from gravimetric analysis. 
As a function of the oxygen deficiency,  the $T_\mathrm{c,on}$ rises rapidly at $\delta\geq \sim 0.2$, then monotonically increases with increasing $\delta$, and finally saturates  around 26-27 K at  $\delta\geq \sim 0.6$.\cite{HA08}
Accordingly, the carriers in the present sample are concentrated  around the optimally doped region. 

The electric resistivity in zero magnetic field was measured by the $dc$ four-terminal method. The magneto-transport up to 9 T was measured by the $ac$ four-probe method using a physical property measuring system (PPMS, Quantum Design), increasing the zero-field-cooling (ZFC) temperatures from 2 K  to 40 K. The high field resistivity  (up to 14 T) was measured in a superconducting magnet at the High Field Laboratory for Superconducting Materials, Institute for Materials Research, Tohoku University.  The electric current  $I$ was applied longitudinally to the sample ; consequently, 
the applied magnetic field $H$ was transverse to the sample (because $H\perp I$). 
Hydrostatic pressures in the electric resistivity measurement were applied using a clamp-type CuBe/NiCrAl cell up to 1.6 GPa. 
Fluorinert was used as a pressure transmitting medium. Inner and outer cylinders were made of NiCrAl and CuBe  alloys, respectively.

%The specific heats in the ZFC mode were measured to be between 2 K and 40 K by the PPMS. 
%For comparison,  the temperature dependence of the specific heat of the superconducting Y$_{2}$Ba$_{4}$Cu$_{7}$O$_{15-\delta }$ (Y247) compound prepared by the high-pressure oxygen method \cite{YA94} was separately measured in zero field at NIMS.  

The $dc$ magnetization was performed under ZFC in a commercial superconducting quantum interference  device magnetometer (Quantum Design, MPMS).
The magnetization measurement under applied pressure was conducted using a standard clamp-type pressure device.
A simple clamp cell consists of  a ZrO$_{2}$ piston and cylinder made of a CuBe alloy. The $M/H $ values of the clamp-type cell are estimated to be $\sim 10^{-8}$
emu/g and are considerably smaller than the values of $M/H$ for all samples. 
% Magnetic fields of 0.002 T, 0.01 T, 1 T, 3 T, and 5 T were applied. 

%\begin{figure}[ht]
%\includegraphics[width=8cm]{FigMT.ps}
%\caption{(color online) Low-temperature dependences of magnetic susceptibilities  $\chi $ of Pr$_{2}$Ba$_{4}$Cu$_{7}$O$_{15-\delta}$ compounds  under various magnetic fields (1 T, 3 T, and 5 T). (a)Non-superconducting as-sintered sample  and (b) superconducting 48-h-reduced sample  with $T_{c,on}=26.5$ K. Inset plots are magnetic data recorded at  10 mT, from which we estimated $T_\mathrm{c}^\mathrm{on}=26.5$ K. Superconducting volume fractions estimated from  present data are also plotted (right vertical axis in inset).  }\label{MT}. 
%\end{figure}

%

\begin{figure}[ht]

\includegraphics[width=8cm]{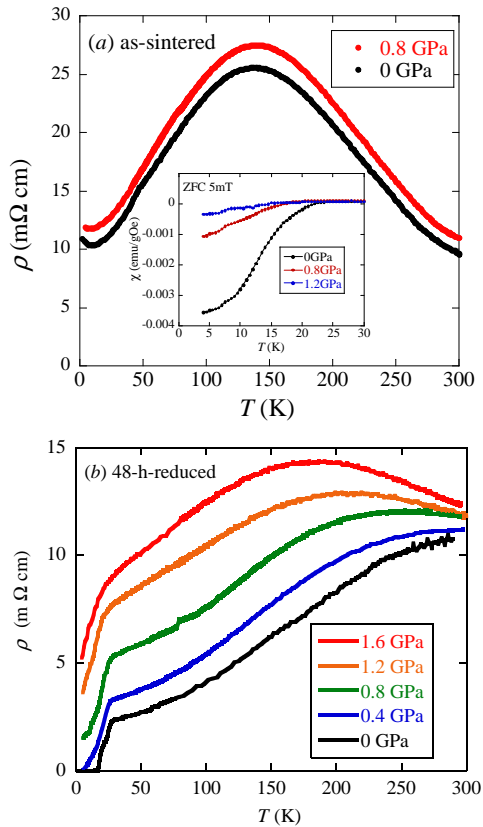}
\caption{(color online) (a) Temperature dependences of electric resistivities of the as-sintered non-superconducting Pr$_{2}$Ba$_{4}$Cu$_{7}$O$_{15-\delta }$ compound measured under 0 GPa and 0.8 GPa. 
(b) Temperature dependences of electric resistivities of the 48-h-reduced superconducting Pr$_{2}$Ba$_{4}$Cu$_{7}$O$_{15-\delta }$ compound
measured under various pressures (0 GPa, 0.4 GPa, 0.8 GPa, 1.2 GPa and 1.6 GPa). Inset  plots low-temperature dependences of magnetic  susceptibilities  $\chi $ of the 48-h-reduced superconducting sample measured under several pressures (up to 1.2 GPa). }
\label{RT}
\end{figure}

\begin{figure}[ht]
\includegraphics[width=8cm]{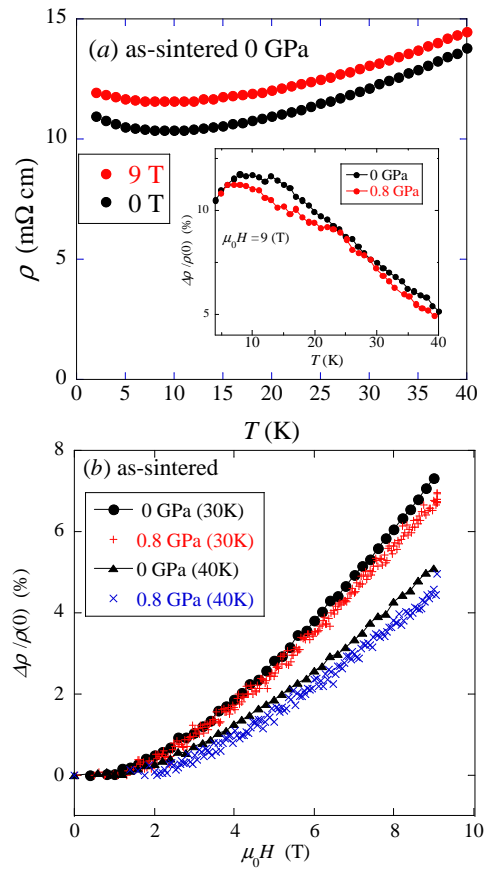}
\caption{(color online) (a) Low-temperature dependences of electric resistivities of the  as-sintered non-superconducting Pr$_{2}$Ba$_{4}$Cu$_{7}$O$_{15-\delta }$ compound measured under a zero field and an applied field of 9 T. Inset plots the magneto-resistance data of the as-sintered sample $\Delta \rho /\rho(0) $ versus $T$ (below 40 K) under 0 GPa and 0.8 GPa. Here, $\Delta \rho = \rho(H)-\rho(0) $.
 In (b),  the magneto-resistance data of the as-sintered sample $\Delta \rho /\rho(0) $  are plotted as a function of $H$ (up to 9 T) at $T=$ 30 K and 40 K  under both $P=$ 0 GPa and 0.8 GPa.}\label{RTAS}. 
\end{figure}

\begin{figure}[ht]
\includegraphics[width=8cm]{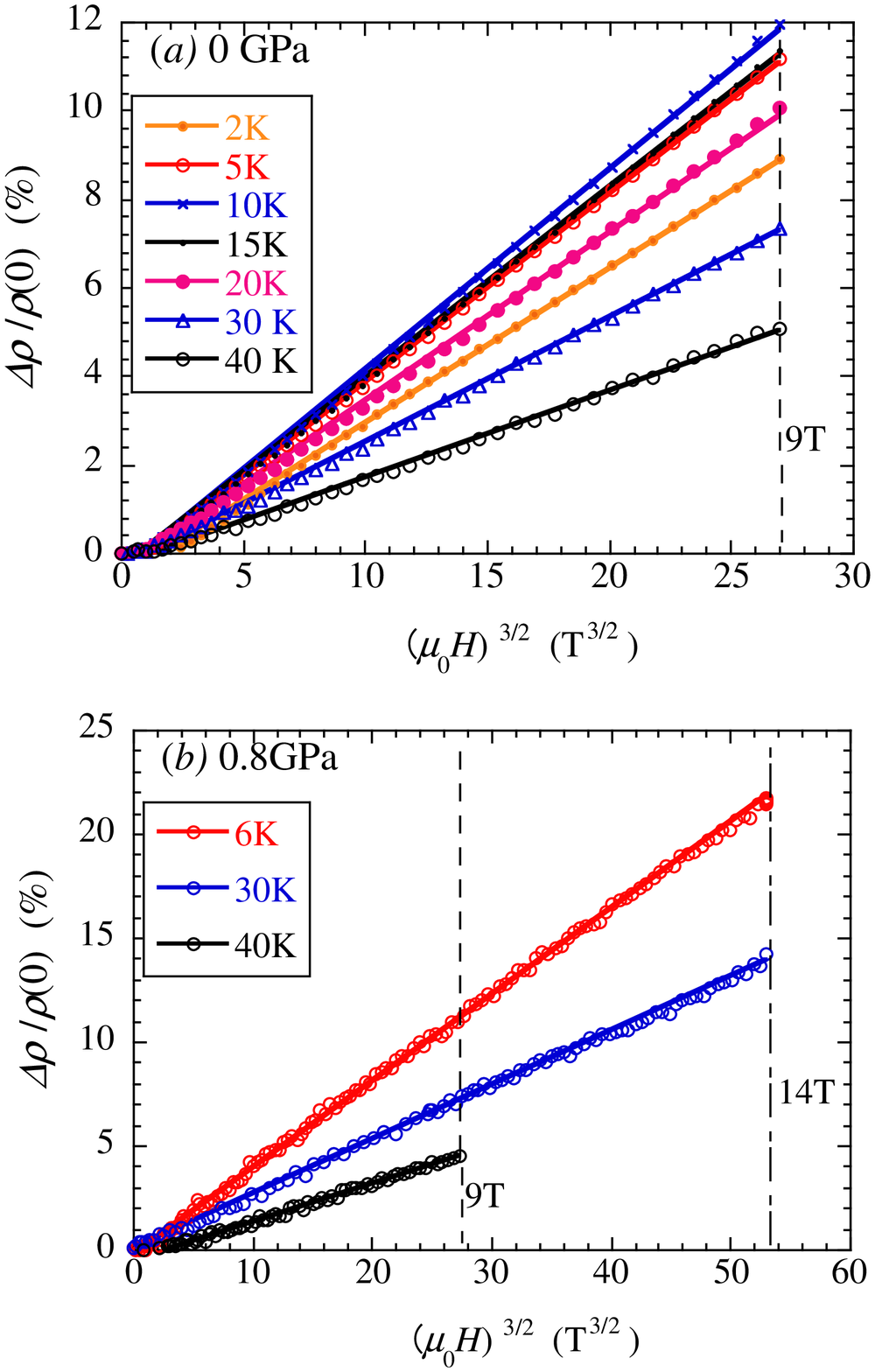}
\caption{(color online) Magneto-resistance  effect  $\Delta \rho /\rho(0) $ of the as-sintered non-superconducting Pr$_{2}$Ba$_{4}$Cu$_{7}$O$_{15-\delta}$ compound. In (a), the MR data (up to 9 T) under ambient pressure are plotted as a function of  $H^{3/2}$ at the selected temperatures ($T$=2 K-40 K).
In (b), the MR data (up to 14 T) under 0.8 GPa are plotted as a function of  $H^{3/2}$ at the selected temperatures of $T$ = 5 K, 30 K, and 40 K.   }
\label{RHAS}. 
\end{figure}

\section{RESULTS AND DISCUSSION}

\begin{figure}[ht]
\includegraphics[width=8cm]{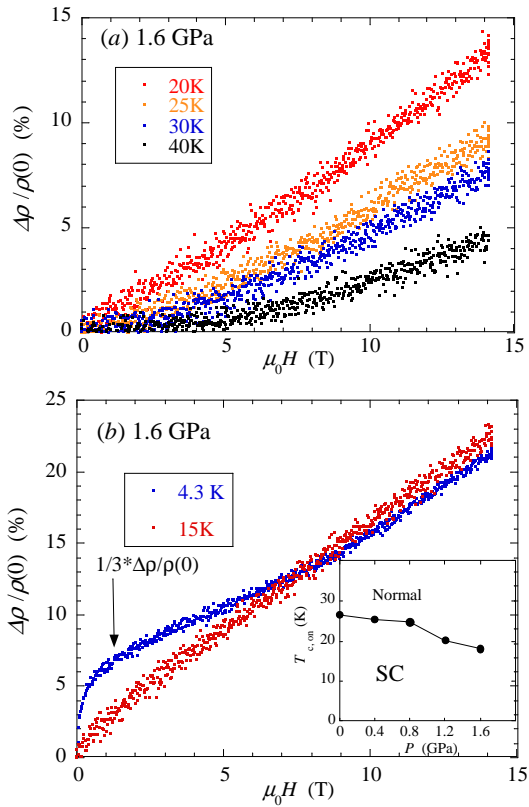}
\caption{(color online) Magneto-resistances (up to 14 T) of the  48-h-reduced  superconducting sample for temperatures close to $ T_{c,on}=18$ K  under a maximum pressure of 1.6 GPa  .   For   $T> T_{c,on}$  and  $T< T_{c,on}$, the MR data are presented in (a) and (b), respectively.   Inset plots $T_{c,on}$ versus pressure (up to 1.6 GPa).}
\label{RHHP}. 
\end{figure}

Figure \ref{RT}  shows the temperature dependences of electric resistivities of the as-sintered non-superconducting and 48-h-reduced superconducting samples measured under various pressures ((a)0.GPa and 0.8 GPa, (b) 0 GPa, 0.4 GPa, 0.8 GPa, 1.2 GPa, and 1.6 GPa). For comparison, inset of Fig.\ref{RT} plots low-temperature dependences of magnetic  susceptibilities  $\chi $ of the superconducting sample under various pressures (up to 1.2 GPa). First of all, the superconductivity of the 48-h-reduced sample was suppressed with increasing the applied pressure.  When applied pressures exceed 0.8 GPa, zero-resistance state of the 48-h-reduced sample then disappeared.  
Second, the high-temperature metallic properties of the reduced superconducting sample were changed to the semi-conducting behaviors at applied pressures above 0.8 GPa, accompanied by a rapid increase in $\rho$.  
The onset temperature of superconducting transition $T_\mathrm{c,on}$ is suppressed from 26.5 K under ambient pressure  through 24.1 K under 0.8 GPa  to 18.0 K under a maximum pressure of 1.6 GPa. ($T_\mathrm{c,on}$ versus pressure at a zero field  as displayed in inset of Fig.\ref{RHHP}) 
%At 1.2 GPa, the superconducting transition onset temperature remains stable, which is close to the inhomogeneous distribution of the superconducting grains.  The superconducting volume fraction was considerably depressed from 30 $\%$ at 0 GPa down to a few percents at 1.2 GPa. 
The  normal-superconducting phase diagram of CuO double chains has been investigated on the basis of the Tomonaga-Luttinger Liquid theory.\cite{SA05}
When the distance between the two single  chains  of a CuO double-chain@block  shrinks along the $c$-axis, the hopping energies $t_{pp}$ and  $t_{dd}$ are expected to increase. 
Here, $t_{pp}$ and  $t_{dd}$ represent the hoping term between  2$p_{\sigma}$ orbitals at the nearest neighbor oxygen sites and that between 3$d_{x^2-y^2}$ orbitals at the nearest neighbor copper sites, respectively.  
The application of the external pressure on the CuO double chains causes an enhancement  of the hopping terms, resulting in a phase transition from the superconducting to normal phase. This theoretical prediction is qualitatively consistent with the superconducting properties suppressed by the pressure effect.\cite{IS07} 

As shown in Fig. \ref{RT}, the pressure effect on the electronic transports of the as-sintered non-superconducting sample is quite weak , in both its magnitude and temperature dependence, in comparison with that of the 48-h-reduced superconducting sample. 
(These behaviors of the present as-sintered Pr247 are almost similar to the effect of pressure on the temperature dependence of as-sintered Pr247 compound prepared by high-pressure synthesis technique. \cite{MA95})

Next, let us show in Fig. \ref{RTAS} the temperature dependences of electric resistivities of the  as-sintered non-superconducting Pr$_{2}$Ba$_{4}$Cu$_{7}$O$_{15-\delta }$ compound measured under a zero field and an applied field of 9 T. 
Inset of Fig.  \ref{RTAS} plots the magneto-resistance (MR) data of the as-sintered sample $\Delta \rho /\rho(0) $ versus $T$ (between 4 K and 40 K) under 0 GPa and 0.8 GPa. 
The value of  $\Delta \rho /\rho(0) $ taken at 9 T reaches a maximum of about 12 $ \%$  near 10 K under either  0 GPa or 0.8 GPa.
Moreover, the magneto-resistance curves as a function of applied field follow similar field dependences in the both cases of 0 GPa and 0.8 GPa. 
Accordingly, we infer  that application of pressure on the as-sintered sample gives rise to no considerable variation in the magneto-resistance effect. 

Here, we try to fit the field dependence of the MR effect $\Delta \rho /\rho(0) $ by using a power law of $H$. 
As shown in Fig.\ref{RHAS}, we found out that $\Delta \rho /\rho(0) $ of the as-sintered Pr247 sample at ambient pressure is well scaled with $ H^{3/2} $ over a wide range of temperatures (2 K-40 K). In a similar way,  the MR data of the as-sintered sample recorded at 0.8 GPa are well described by  a power law of $\Delta \rho /\rho(0) \propto H^{3/2} $.
Our findings are not consistent with the  MR effect of polycrystalline Pr124 compounds with metallic double-chains. 
In Pr124 system, $\Delta \rho /\rho(0) \propto H^{2} $ for $H\perp I$.\cite{TE96}
These discrepancies in magnetic field dependences of MR between Pr124 and Pr247 are originated from the differences of their crystal structures.
In the latter compounds, we believe that CuO double chains stacked along the c-axis are separated  by CuO singe chains, resulting in a suppression of carrier hopping between individual double chains along the c-axis.  
Furthermore, a photoemission spectroscopy study\cite{WA08} has revealed that the CuO double chain in Pr247 has a better one dimensionality than that in Pr124, which is probably close to the suppression of c-axis carrier hopping. 

In Pr124 system,  it is well known that the inter-chain transport at low temperatures is destroyed by a high magnetic field applied perpendicular to the CuO double chains.\cite{HU02}
The field-induced dimensional-crossover causes the weak field dependence of the c-axis magneto-resistance of Pr124 when the applied field exceeds the decoupling field of 11 T.
Furthermore, in Y124 system,   the field dependence of the c-axis magneto-resistance is changed from $ H^{2} $ to $H$-linear dependence at high fields, suggesting
the field-induced dimensional crossover.\cite{HU98} An anomalous field dependence of the $c$-axis MR ( $\Delta \rho \propto H^{3/2} $)  has been reported in the 1D organic conductor  
(TMTSF)$_{2}$PF$_{6}$ once the inter-chain transport becomes incoherent.\cite{DA95}
For Pr124 or Y124 with \{-D-D-D-D-\} sequence, the c-axis hopping of carriers between CuO double chains is suppressed by the application of highly magnetic field.
In Pr247 system, the c-axis carrier transport is originally incoherent because of regular stacking structures of \{-D-S-D-S-\} along the $c$-axis direction, as depicted in Fig.\ref{Xray}.

%In the presence of  hydrostatic pressure (0.8 GPa), the  magneto-resistances taken at 30 K  show a slightly decrease by  $\sim 0.5 \%$ in the comparison to 
%those values without pressures.  These pressure insensitive trends considerably differ from the strong pressure dependences of MR effect for the superconducting Pr247. 

%In the as-sintered non-superconducting sample,  the magneto-resistances of Fig.\ref{RHAS}  follow a power law with exponent 1.6-1.7 at temperatures below 40 K.

%In a previous study\cite{WA12},  Wakeham et al. assumed that the typical Fermi surface of a 1D conductor is not straight but warped, in order to explain temperature dependence of Hall coefficient of Pr124.  
%The MR effect of the as-sintered non-superconducting Pr247 is,  in its temperature dependence and magnitude, similar to that of the polycrystalline Pr124.
%For Pr124, $\Delta \rho /\rho(0) $ at 9 T reaches at least $\sim 10\%$ at 30 K, which is comparable to the MR value of the as-sintered Pr247.\cite{TE96}

\begin{figure}[ht]
\includegraphics[width=8cm]{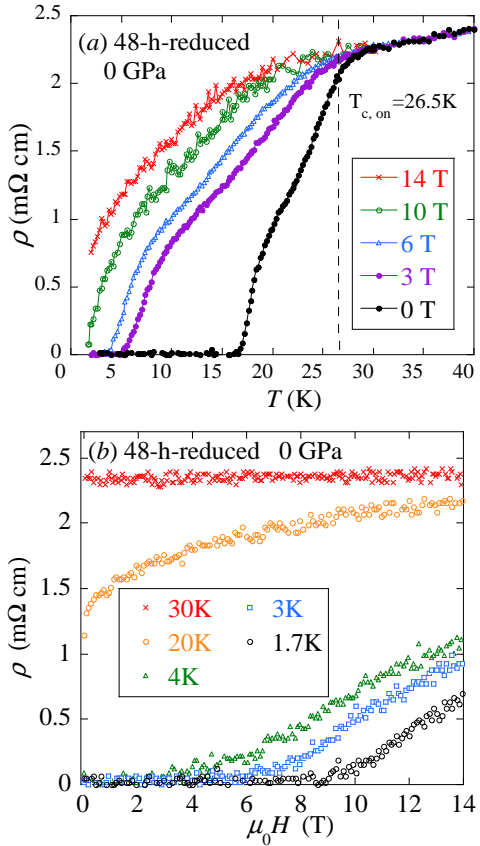}
\caption{(color online) (a) Low-temperature dependences of electric resistivities of the 48-h-reduced  superconducting Pr$_{2}$Ba$_{4}$Cu$_{7}$O$_{15-\delta }$ compound measured at the several applied fields ($H$ = 0 T-14 T).
 In (b),  the magneto-resistance data of the  superconducting sample are plotted as a function of magnetic fields (up to 14 T) at the selected  temperatures ($T=$ 1.7 K-30 K). }\label{RTH}. 
\end{figure}

\begin{figure}[ht]
\includegraphics[width=10cm]{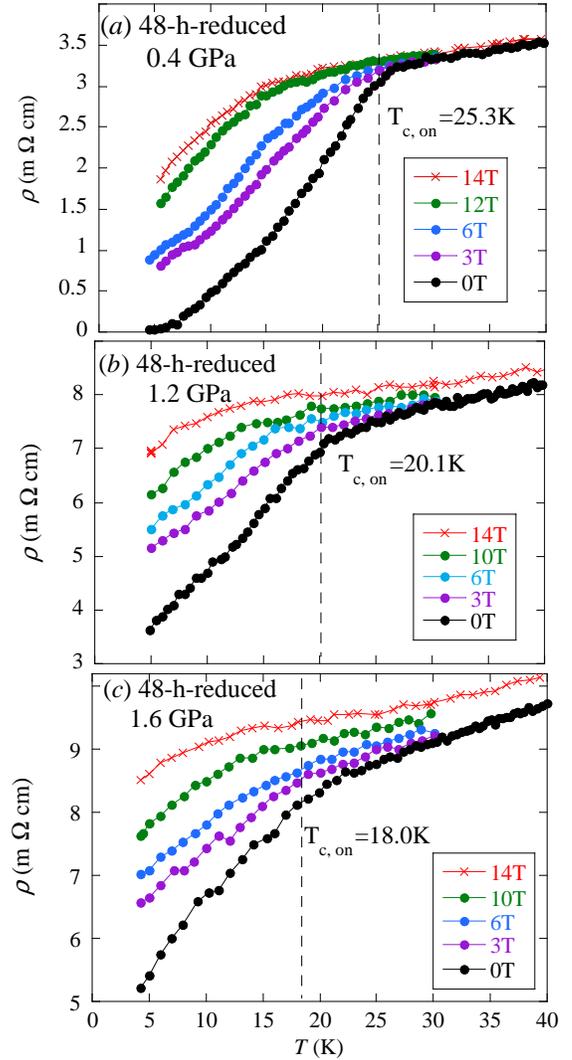}
\caption{(color online) Low-temperature dependences of electric resistivities of the 48-h-reduced  superconducting Pr$_{2}$Ba$_{4}$Cu$_{7}$O$_{15-\delta }$ compound measured at several applied fields  under hydrostatic pressures of (a)0.4 GPa, (b) 1.2 GPa, and (c) 1.6 GPa. For $H$ = 0 Tesla, $T_\mathrm{c,on}$ is decreased from 25.3 K at 0.4 GPa  through 20.1 K under 1.2 GPa  to 18.0 K under a maximum pressure of 1.6 GPa.
  }
\label{RTP}. 
\end{figure}

\begin{figure}[ht]
\includegraphics[width=8cm]{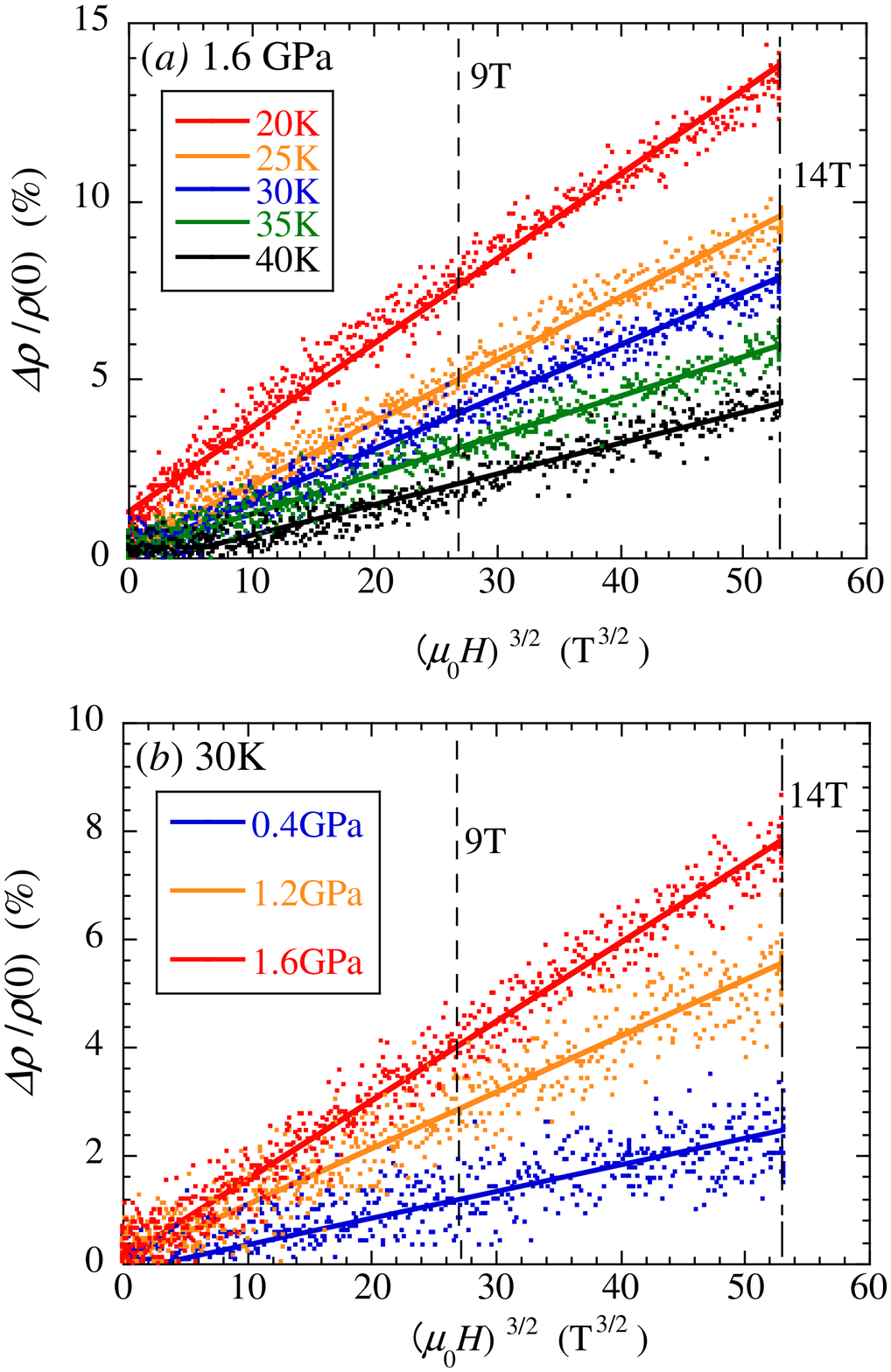}
\caption{(color online)The corresponding MR data (up to 14 T) are plotted as a function of  $H^{3/2}$.
The data of (a) and (b) are collected under $P$ = 1.6 GPa at  $T$ = 20 K, 25 K, 30 K, 35 K, and 40 K, and at  $T$ = 30 K under $P$ = 0.4 GPa, 1.2 GPa, and 1.6 GPa, respectively. }
\label{RHHP2}. 
\end{figure}

Figure \ref{RTH} (a) plots the low-temperature  dependences of electric resistivities of the superconducting Pr$_{2}$Ba$_{4}$Cu$_{7}$O$_{15-\delta}$ compound with $T_{c,on}=26.5$ K in several applied fields (0 T, 3 T, 6 T, 10 T, and 14 T). Even at high fields above 10 T, the resistive drop associated with the superconducting transition is observed but the diamagnetic signal is strongly  suppressed at low field below 1 T.  This observation is closely related to the 1D superconductivity of CuO double-chains as highlighted in our previous paper. \cite{CH13}
The magneto-resistance effect (up to 14 T) of the 48-h-reduced superconducting sample in the normal sate temperature region above $T_{c,on}=26.5$ K is very small in comparison to the large MR effect of the as-sintered sample.  $\Delta \rho /\rho$ of the as-sintered sample recorded at 30 K  attains $\sim 7\%$ up to 9 T while for the 48-h-reduced sample the MR effect rapidly decreases below $\sim 1\%$ at 9 T.  The suppressed MR effect in the normal phase of  the 48-h-reduced superconducting sample is probably explained by the topology of Fermi surfaces of CuO double chains\cite{NA07}, where the Fermi surfaces are changed from their warped to straight forms upon increasing  carrier contents. (as mentioned below)

%In a previous study\cite{WA12},  Wakeham et al. assumed that the typical Fermi surface of a 1D conductor is not straight but warped, in order to explain temperature dependence of Hall coefficient of Pr124.  

%The magnetoresistance of the as-sintered and 48-h-reduced samples  differ at intermediate temperatures (above $T_\mathrm{c}^\mathrm{on}=26.5$ K).

Figure \ref{RTP}  plots the low-temperature dependences of electric resistivities of the superconducting Pr$_{2}$Ba$_{4}$Cu$_{7}$O$_{15-\delta}$ compound measured in the several applied fields (0 T-14 T) under various hydrostatic pressures ((a)$P$ = 0.4 GPa, (b)$P$ = 1.2 GPa, and (c)$P$ =1.6 GPa). 
For the superconducting sample without magnetic field, the superconducting transition temperature reaching zero-resistance state  $T_{c,zero}$ is suppressed from 16 K at 0 GPa down to 6 K at a low pressure of 0.4 GPa. 
At high fields above 10 T, the low-temperature  data under 0.4 GPa remain  the resistive drops associated with the superconducting transition, upon decreasing temperature.
These tendencies still appear for the  high-field resistive data (up to 10 T) under both 1.2 GPa and 1.6 GPa. 

%However, under both conditions of 14 T and  1.2 GPa  the resistive decrease due to the superconductivity is no visible.  Both effects of moderate pressure and applied high field on the 48-h-reduced sample completely suppresses the superconductivity of metallic CuO double chains

For temperatures close to  $T_{c,on}$,  we are concentrated on the magneto-transport properties in the normal state of the 48-h-reduced superconducting sample.
As mentioned before, although the strong MR effect  is observed in  the as-sintered non-superconducting sample, the 48-h-reduced annealing procedure for it  causes a considerably suppression of the normal-state MR accompanied by the appearance of the superconductivity with $T_{c,on}= 26.5$ K. 
The application of pressure on the superconducting sample gives rise to a substantial decrease of  $T_{c,on}$  from 26.5 K at 0 GPa  down to  $T_{c,on}= 18$ K at 1.6 GPa, as shown in inset of Fig. \ref{RHHP}. 
On the other hand,  the finite magneto-resistance effect reappears for $T\geq T_{c,on}$  when the applied pressure to the superconducting sample is increased. 
%We note that the high-field resistivities  ($H\geq $ 10 T)  at the normal-state temperatures ($T\geq T_{c,on}$ ) exhibit larger MR effect,  as the applied pressure is increased from  0.4 GPa  through 1.2 GPa up to 1.6 GPa, as shown in Fig. \ref{RTP}. 

Figure \ref{RHHP} plots  the magneto-resistances (up to 14 T) of the  48-h-reduced  superconducting sample, $\Delta \rho /\rho(0) $, measured under a maximum pressure of 1.6 GPa.  For $T > T_{c,on}=18$ K,  $\Delta \rho /\rho(0) $ curves  initially start to increase in the  the  downward convey forms with applied fields.
On the other hand,  the MR data recorded at $T$ below  $T_{c,on}= 18$ K tend to increase  according to the upward covey behaviors at low fields. 

In a similar way to the MR of the as-sintered sample,  the MR data of the superconducting sample recorded under 1.6 GPa are well described by  a power law of $\Delta \rho /\rho(0) \propto H^{3/2} $ over the normal state region between 20 K and 40 K, as shown in Fig.\ref{RHHP2} (a).   
For comparison, Fig.\ref{RHHP2} (b) shows the MR data at  $T$ = 30 K under $P$ = 0.4 GPa, 1.2 GPa, and 1.6 GPa.   The application of pressures on the superconducting sample enhances  the MR at 14 T from 2.2 $\%$ at 0.4 GPa through 5.8 $\%$ at 1.2 GPa up to 8 $\%$ at 1.6 GPa. Furthermore, the MR data  under pressures are well fitted by using  $\Delta \rho /\rho(0) \propto H^{3/2} $.(Fig. \ref{RHHP2}(b))

%In (a), the MR data are presented at $T$ = 20 K, 25 K, 30 K, and 40 K ($> T_{c,on}=\sim 18$ K). For comparison, in (b) the MR data  are recorded at $T$ = 4.3 K and 15 K ($< T_{c,on}=\sim 18$ K). Inset plots $T_{c,on}$ versus pressure (up to 1.6 GPa).
%For comparison, the MR data of the as-sintered non-superconducting sample were given.
%At the intermediate temperature of 30 K,  $\Delta \rho /\rho(0) $ of the 48-h-reduced sample at the maximum field of 14 T  is increased to about 2.7 $\%$ under a low pressure of 0.4 GPa,  then it shows a further increase up to $\sim $5.8 $\%$ under 1.2 GPa, and it  finally reaches $ \sim $7.4 $\%$ under a maximum pressure of 1.6 GPa. 

%Here, we notice that the application of external pressure on the 48-h-reduced sample induces the realization of clear magneto-resistance effect.
%ssuming that $\Delta \rho /\rho(0) \propto H^{n} $, the exponent values at 30 K are roughly estimated from the MR data of (a)  to be 1.1 at 0.4 GPa, 1.1 at 0.8 GPa , and 1.2 at 1.2 GPa. As the applied pressure is increased, the MR effect of the superconducting sample is not only enhanced but also  the exponent value of the power law is made larger.
%To our knowledge, these pressure-induced MR effects observed in the normal state of superconducting materials are probably  novel phenomena  and are understood on the basis of Fermi surface topology as mentioned below. 

Now, we try to discuss the  pressure-induced magneto-resistance effect of the 48-h-reduced superconducting sample.
First of all,  we assume that the Fermi surfaces of the double chains of the superconducting Pr247 are almost straight as indicated by the band calculation.\cite{HA11} 
%Here, $k_{x}$ and $k_{y}$  denote the wave number components  along and across the CuO  double-chain direction, respectively, 
In an ideal 1D conductor with straight Fermi surface without the wave number $k_{y}$,  the orbitals of itinerant carries across the CuO double chains are not bended by the Lorentz force, resulting in no MR effect.
In the case of warped Fermi surface,  the itinerant carriers have the finite component of wave number across the CuO double chain direction and the Lorentz force can bend their orbitals, giving the classical MR effect.  For the superconducting Pr247, CuO double chains have straight Fermi surfaces at ambient pressure, which is consistent with the absence of  MR effect observed.
On the other hand,  if the Fermi surfaces of double chains under the external pressures  are changed to their warped structures with finite values of  $k_{y}$.
then it is naturally understood that the MR effects appear with increasing the applied pressure. 
In particular,  the lattice shrinkage due to the hydrostatic pressure enhances not only charge transfer across two single chains of a CuO double chain  
but also increase the couplings between  CuO double chains along the a-axis. 
As mentioned before, the suppressed  MR effect of the superconducting sample under ambient pressure  is closely related to the topology of Fermi surfaces of CuO double chains, which  are varied from their warped to straight forms upon increasing  carrier contents.   
Accordingly,  we expect that the model of warped Fermi surfaces is not only applicable to the  origins of the MR effect of the  non-superconducting sample, but is also related to the reasons for the pressure-induced MR phenomena of the superconducting sample. 

%It is true that  the  oxygen-removal process gives rise to electron doping into the CuO double-chain blocks, resulting in the superconductivity  with $T_{c,on}=26.5$ K. However, 
%it has not been made clear the reason why  the magneto-resistance effect of  the 48-h reduced sample  vanishes  in the normal-state temperature region, in spite of the appearance of the superconducting phase.
%The oxygen removal treatment on the as-sintered  Pr$_{2}$Ba$_{4}$Cu$_{7}$O$_{15-\delta }$  elongates the $a$-axis across the CuO double-chain direction from 3.89 at $\delta=0$  to 3.90 
%at $\delta=\sim 0.5$. This slight  increase in the lattice parameter  probably causes a variation of the inter-chain hopping parameter  between double chains along the  $a$-axis.
%
% 

\section{SUMMARY}
  
We demonstrated the magneto-transport properties of superconducting and non-superconducting Pr$_{2}$Ba$_{4}$Cu$_{7}$O$_{15-\delta }$ compounds under the application of different hydrostatic pressures (up to 1.6 GPa). 
We found out that $\Delta \rho /\rho(0) $ of the as-sintered Pr247 sample at ambient pressure is well scaled by using  $ H^{3/2} $ over a wide range of temperatures (2 K-40 K). 
In present Pr247 system,  crystal structures of \{-D-S-D-S-\} along the $c$-axis direction probably make inter-chain couplings between CuO double chains incoherent, resulting in the  $ H^{3/2} $ dependence of MR.  This result of Pr247  is not consistent with the $ H^{2} $ dependence of MR in  Pr124 system with stacking structures of \{-D-D-D-D-\} along the $c$-axis. 
The superconducting properties suppressed by the pressure effect is qualitatively explained by 
the  normal-superconducting phase diagram of CuO double chains on the basis of the Tomonaga-Luttinger Liquid theory. 
The 48-h-reduced annealing procedure for the as-sintered sample   causes a considerably decrease in the normal-state MR accompanied by the appearance of the superconductivity with $T_{c,on}= 26.5$ K. 
The suppressed  MR effect of the superconducting sample is probably explained by the topology of Fermi surfaces of CuO double chains, where the Fermi surfaces are changed from their warped to straight forms upon increasing  carrier contents.   
In contrast, for  $T > T_{c,on}$, the application of pressure (up to 1.6 GPa) on the superconducting sample results in a substantial MR effect of $8 \%$ at 14 T.  
In a similar way to the MR of the as-sintered sample,  the MR data of the superconducting sample recorded under 1.6 GPa are well described by  a power law of $\Delta \rho /\rho(0) \propto H^{3/2} $ over the normal state region between 20 K and 40 K.
The model of slightly warped Fermi surfaces explains  not only the magneto-resistance effect of the non-superconducting sample, but it also is closely related to the reasons for  the pressure-induced MR phenomena of the superconducting sample. 

%X-ray powder diffraction pattern of nominal Pr$_{2}$Ba$_{4}$Cu$_{7}$O$_{15-\delta }$ reveals the coexistence of Pr123 and Pr124 phases.  
\begin{acknowledgments}
The authors are grateful for  M. Nakamura for his assistance in PPMS experiments at Center for Regional Collaboration in Research and Education, Iwate University. 
%They thank Dr. A. Matsushita for his collaboration in Hall coefficient measurement. 
%This work was partially supported by a Grant-in-Aid for Scientific Research from Japan Society of the Promotion of Science. 
\end{acknowledgments}

\end{document}